\documentclass[12pt]{iopart} 
\usepackage{amssymb} \usepackage{graphicx} \usepackage{float}
\usepackage{color}

\begin{document}

\title{Annealing cycles and the self-organization of functionalized colloids}

 \author{Crist\'{o}v\~{a}o S. Dias, Nuno A. M. Ara\'{u}jo, and Margarida M. Telo da Gama}

\address{Centro de F\'isica Te\'orica e Computacional, Faculdade de
Ci\^encias, Universidade de Lisboa, 1749-016 Lisboa, Portugal \\ Departamento de
F\'isica, Faculdade de Ci\^encias, Universidade de Lisboa, 1749-016 Lisboa, Portugal}
\ead{csdias@fc.ul.pt} \vspace{10pt} \begin{indented} \item[]September 2017
\end{indented}

\begin{abstract}
The self-assembly of functionalized (patchy) particles with directional interactions into target structures 
is still a challenge, despite the significant experimental advances on their synthesis.
The self-assembly pathways are typically characterized by high energy barriers that hinder 
the access to stable (equilibrium) structures. A possible strategy to tackle this challenge is to perform annealing cycles. 
By periodically switching on and off the inter-particle bonds, one expects to smooth-out the 
kinetic pathways and favor the assembly of targeted structures. Preliminary results have shown that the 
efficiency of annealing cycles depends strongly on their frequency. Here, we study numerically how this 
frequency-dependence scales with the strength of the directional interactions (size of the patch $\sigma$). 
We use analytical arguments to show that the scaling results from the statistics of a random walk in configurational space.

\end{abstract}

  \maketitle

\section{Introduction}

Patchy (colloidal) particles are functionalized particles with directional interactions. Studies of thermodynamic phase diagrams predict that, under equilibrium conditions, 
these particles self-assemble into low-density gel-like structures \cite{Dobnikar2013,Zaccarelli2007,Elliott2003,Ruzicka2011,Sciortino2004}. 
However, these low-density structures should have simultaneously mechanical 
stability (for practical applications) and reversibility (to be accessible), posing several challenges to the use of self-assembly techniques 
\cite{Dias2013,Dias2016,Chakrabarti2014,Zaccarelli2006}.

Various methods have been developed to functionalize colloidal particles, e.g., coating their surface with metal, polymers, or DNA 
\cite{Shum2010,Duguet2011,Sacanna2011,He2012,Hu2012,Wilner2012}. 
A popular choice is the use of DNA, which allows for selective and fine-tuned pairwise interactions 
\cite{Kern2003,Frenkel2011,Reinhardt2014,Angioletti-Uberti2014,DiMichele2014}. However, with DNA the typical 
bond energy is of the order of the thermal energy (or even higher) and thus the bonds are typically irreversible within the relevant timescales, 
leading to kinetically arrested structures ~\cite{Geerts2010,Wang2012,Leunissen2011,Joshi2016}. A strategy to overcome the 
energy barriers is to perform annealing cycles \cite{Grzelczak2010,Malinge2016}. The sharp change of the 
DNA bonding probability around the melting temperature, opens the possibility of switching on and off the bonds at will, 
by adjusting the temperature \cite{DiMichele2014,Geerts2010,Mirkin1996,Nykypanchuk2008}. 
Other strategies for equivalent systems have been devised not only using temperature but also light \cite{Bergen2016}.

Recent numerical results show that the efficiency of annealing cycles is maximized at an optimal frequency \cite{Araujo2017,Dias2017}. There, it is hypothesized that 
this non-monotonic dependence results from the competition between two mechanisms: rotational and translational diffusion. When the bonds are switched off, 
particles diffuse freely until the bonds are switched on again. For sufficiently high frequencies, the short off-time is not enough for particles to rotate 
relative to each other and effectively break their bonds. For sufficiently low frequencies, the off-time is so long that, 
at the end of a cycle not only the relative (rotational) orientation 
is decorrelated but also the translational positions are significantly different. Thus, an intermediate time is required to guarantee that only a fraction of the 
bonds are restructured in each cycle.

Here we investigate numerically how the optimal frequency scales with the model parameters. In particular, we are interested on its dependence 
on the width of the interaction range (size of the patch) and diffusion properties. We develop analytical arguments to show that the observed scaling 
results from the statistics of a random walk. The paper is organized as follows. In the next section, we present the model and the details of the simulations. 
The numerical results are discussed in Sec.~\ref{sec::anneal} and the analytical arguments developed in Sec.~\ref{sec::timescales}. In Sec.~\ref{sec::final}, 
we draw some conclusions.


\section{Model and Simulations}\label{sec::model}
\begin{figure}
\begin{center}
\includegraphics[width=0.7\columnwidth]{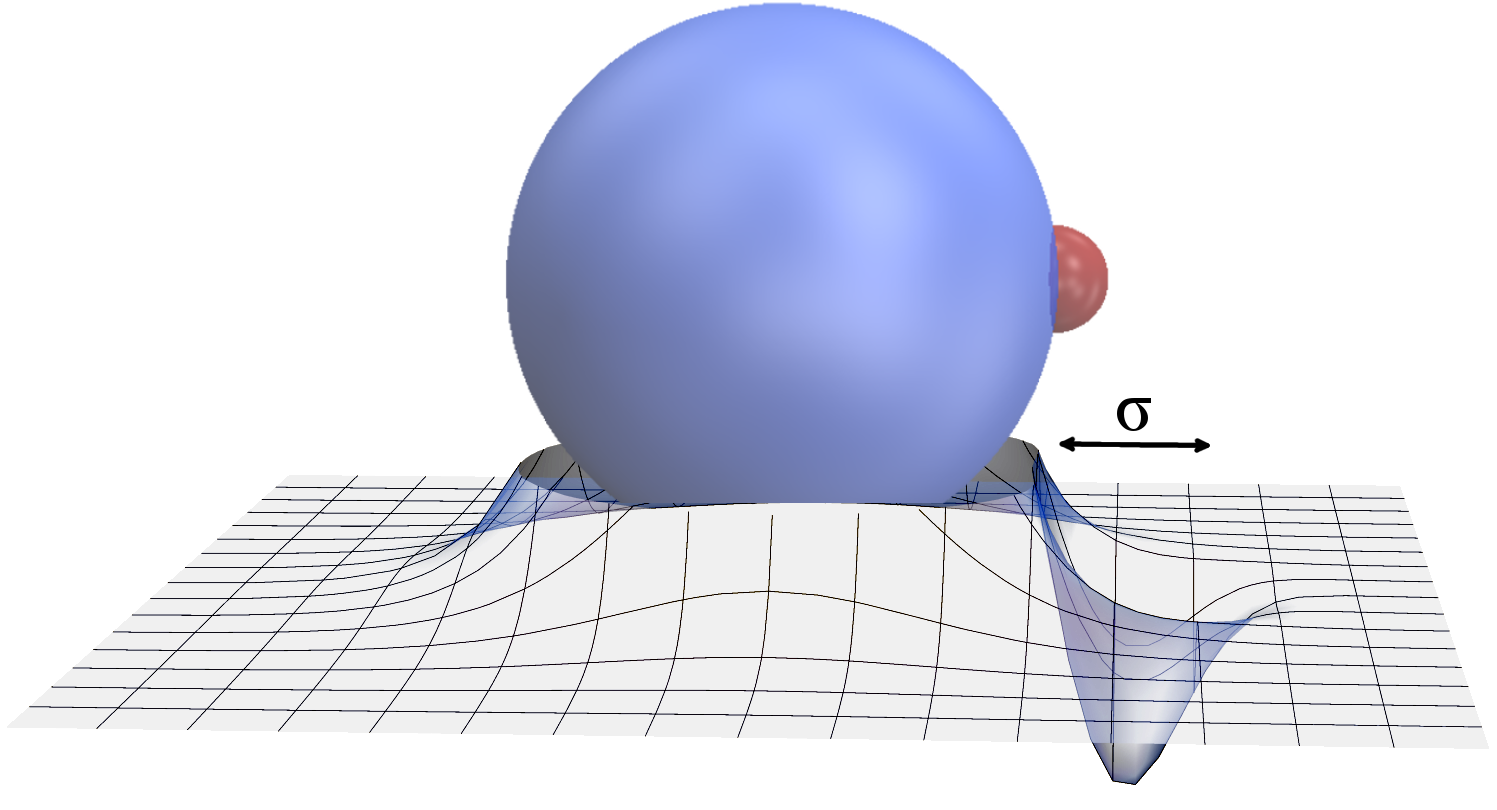}\\
\caption{Schematic representation of a spherical particle (blue) with an attractive patch (red)
on its surface. The energy landscape represents the interaction energy of a probe patch with
the patchy particle, where the interaction with the core is repulsive and the patch-patch 
interaction is attractive. $\sigma$ is the size of the patch, defined as the width of the attractive interaction potential. \label{fig.model}}
\end{center}
\end{figure}

We consider a monodisperse suspension of spherical particles with three, equally-distributed patches, on their surface. 
The interparticle core-core interaction is described by a 
Yukawa-like potential, $V_Y(r)=\frac{A}{k}\exp{\left(-k\left[r-2R\right]\right)}$, 
where $R$ is the effective radius of the particles, $A$ is the
interaction strength and $k$ the inverse of the screening length. The patch-patch attractive interaction between the patches
\cite{Vasilyev2013} (see Fig.~\ref{fig.model}) is described by the pairwise potential $V_G(r_p)=-\epsilon\exp\left[(r_p/\sigma)^2\right]$,
where $\epsilon$ is the interaction strength, $\sigma=\{0.05, 0.075, 0.1, 0.125\}R$ the size of the patch, and $r_p$ the 
distance between patches. 
The resulting energy landscape for a probing patch is schematized in Fig.~\ref{fig.model}.
The parameters are chosen such that the bonds are irreversible within the simulation timescale.
We consider an attractive substrate, interacting isotropically with the particles, with an interaction 
derived from the Hamaker theory for two interacting spherical particles in the limit where the radius of one of them diverges \cite{Everaers2003}.
This results in the superposition of two contributions given by an attractive term,
\begin{equation}
V_A=-\frac{A_H}{6}\left[\frac{2R(R+D)}{D(D+2R)}+\ln\left(\frac{D}{D+2R}\right)\right], \label{eq.Hamaker_att}
\end{equation}
and a repulsive one,
\begin{equation}
V_R=\frac{A_H\sigma^6}{7560}\left[\frac{6R-D}{D^7}+\frac{D+8R}{(D+2R)^7}\right], \label{eq.Hamaker_rep}
\end{equation}
where $A_H$ is the Hamaker's constant and $D=r-R$ the distance between the 
surface of the particle and the substrate. 

We perform Molecular Dynamics simulations, where we integrate the Langevin 
equations of motion for the translational and rotational degrees of freedom given by,
\begin{equation}
 m\dot{\vec{v}}(t)=-\nabla_{\vec{r}} V(\vec{r})-\frac{m}{\tau_t}\vec{v}(t)+\sqrt{\frac{2mk_BT}{\tau_t}}\vec{\xi_t}(t), \label{eq.Langevin_dynamics_trans} 
\end{equation}
and
\begin{equation}
 I\dot{\vec{\omega}}(t)=-\nabla_{\vec{\theta}} V(\vec{\theta})-\frac{I}{\tau_r}\vec{\omega}(t)+\sqrt{\frac{2Ik_BT}{\tau_r}}\vec{\xi_r}(t), \label{eq.Langevin_dynamics_rot}
\end{equation}
respectively, where, $\vec{v}$ and $\vec{\omega}$ are the translational and angular velocity, $m$ and $I$ are
mass and moment of inertia of the patchy particle, $V$ is the pairwise potential, $\tau_t$ and $\tau_r$ are the 
translational and rotational damping times, and $\vec{\xi_t}(t)$ and $\vec{\xi_r}(t)$ are stochastic terms
taken from a random distribution of zero mean. We use the velocity
Verlet scheme and the Large-scale Atomic/Molecular Massively Parallel Simulator (LAMMPS) \cite{Plimpton1995}. We 
consider the damping time for the translational motion, $\tau_t=m/(6\pi\eta R)$, which from the Stokes-Einstein-Debye
relation \cite{Mazza2007},
\begin{equation}
 \frac{D_r}{D_t}=\frac{3}{4R^2}, \label{eq.coeff_rel}
\end{equation}
gives a relation for the damping time of the rotational degrees of freedom $\tau_r=10\tau_t/3$.

As in Ref.~\cite{Dias2016,Araujo2017}, the initial configurations were generated using the three-dimensional 
ballistic deposition model described in Refs.~\cite{Dias2013,Dias2013b}. In all cases, we start with a 
coverage of $\pi/\left[3\tan\left(\pi/3\right)\right]$, which corresponds to the one of the honeycomb lattice.

\section{Annealing cycles}\label{sec::anneal}

\begin{figure}
\begin{center}
\includegraphics[width=0.8\columnwidth]{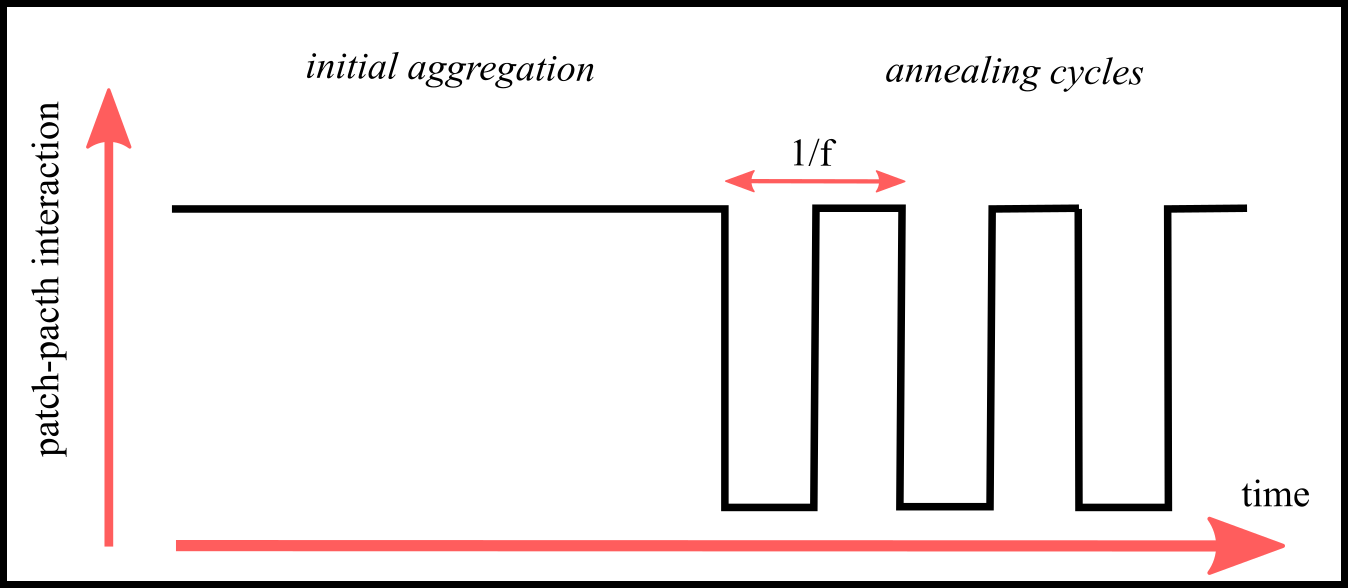}\\
\caption{Schematic representation of the annealing protocol. The 
patch-patch interaction is switched on during the initial aggregation time, where an irreversible structure is formed.
After the initial aggregation time, the annealing cycle is initiated, where the patch-patch interaction is sequentially switched on
and off at a certain frequency $f$.\label{fig.annealing_model}}
\end{center}
\end{figure}

To study the annealing cycles, we considered the following protocol (see also scheme in Fig.~\ref{fig.annealing_model}).  Particles adsorbed on the substrate
are let to evolve according to the equations of motion \ref{eq.Langevin_dynamics_trans} and \ref{eq.Langevin_dynamics_rot} during $10^3\tau_B$ ($\tau_B$ 
is the Brownian time, i.e., the time a single particle
takes to diffuse a area of $(2R)^2$, where $R$ is the effective radius of the particle). For the considered range of parameters, 
during this initial time, the particles diffuse and aggregate irreversibly. 
After that, an annealing cycle is performed for an
additional time of $2000\tau_B$. The attractive
interaction is sequentially switched on and off for time intervals of $f^{-1}$, where $f$ is the frequency of the 
annealing cycle in units of $\tau_B^{-1}$.

\begin{figure}
\begin{center}
\includegraphics[width=0.8\columnwidth]{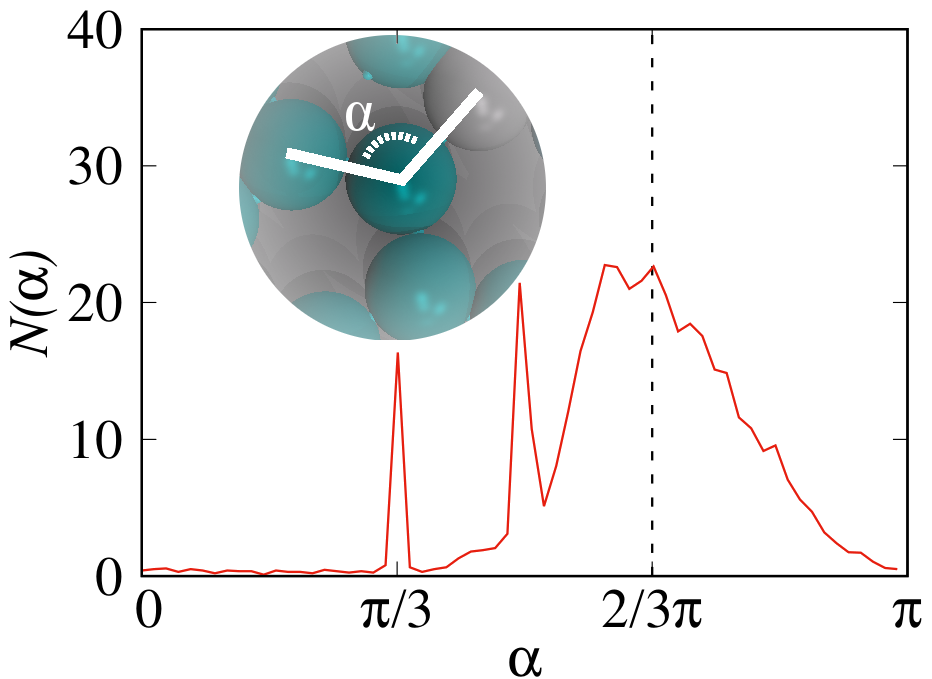}\\
\caption{Angular distribution function, $N(\alpha)$, where $\alpha$ is the angle 
formed by two bonds on a connected patchy particle
with $\sigma=1$, in a system of linear size $L=32$, averaged over 20 samples, and a coverage of $\pi/\left[3\tan\left(\pi/3\right)\right]$. 
The peaks represent the angles observed more frequently. The dashed line is the expected angle for a
honeycomb lattice (for three equally distributed bonds). Inset: Schematic representation of the
measurement of the angle $\alpha$ between bonds. \label{fig.angle_distribution}}
\end{center}
\end{figure}

To evaluate the structure, we measure the angular distribution function
for the patch-patch bond orientation, given by,
\begin{equation}\label{eq::angdistrib}
N(\alpha)=\sum^{N_p}_{n=1}\left[\sum_{i}^{bonds-1}\sum_{j>i}^{bonds} g\left(\alpha_{ij}^n-\alpha^n,\delta \alpha\right)\right] ,
\end{equation}
where $N_p$ is the total number of particles, $bonds$ the number of bonds in particle $n$, $\alpha_{ij}^n$ is the angle
between bonds $i$ and $j$ of particle $n$ (see inset of Fig.~\ref{fig.angle_distribution}), and the function $g\left(\alpha_{ij}^n-\alpha^n,\delta \alpha\right)$ is either one, if 
$\left|\alpha_{ij}^n-\alpha^n\right|<\delta \alpha$, or zero ($\delta \alpha=0.01$ rad). The characteristic peak for the honeycomb lattice is
at $\alpha=2\pi/3$. The angular distribution function, after the 
initial aggregation, is shown in  Fig.~\ref{fig.angle_distribution}. A very wide maximum is observed around the characteristic peak of the honeycomb lattice, 
together with two additional peaks at $\alpha=\pi/3$ and $\alpha=\pi/2$, due to three- and four-particle aggregates. 
This result indicates that the structure differs significantly from the ordered (honeycomb) structure expected at this density and distribution of patches.

\begin{figure}
\begin{center}
\includegraphics[width=0.8\columnwidth]{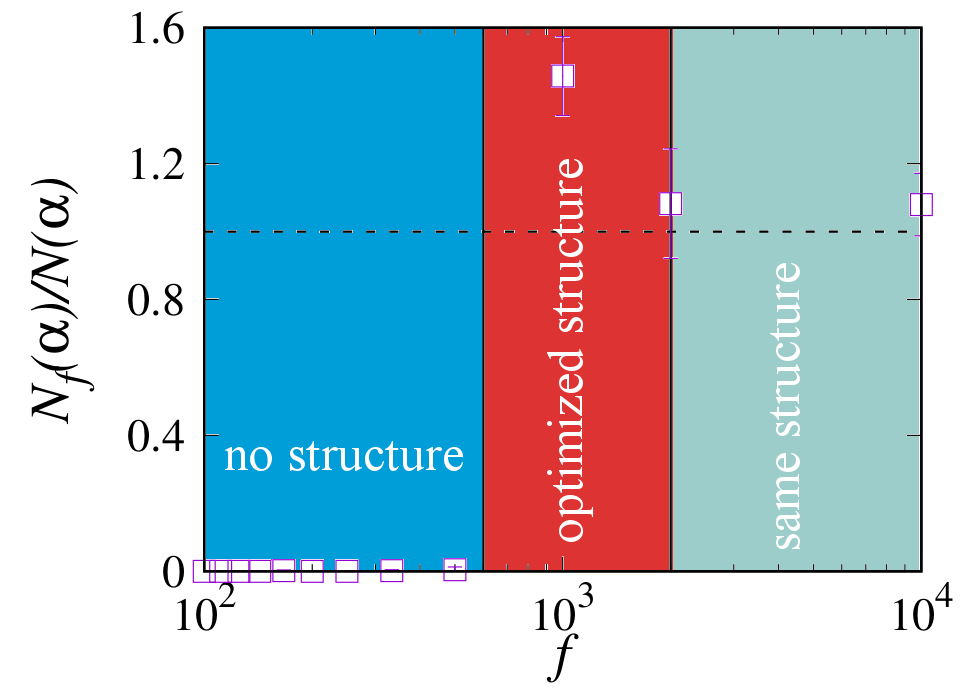}\\
\caption{Value of the angular distribution function $N(\alpha)$ for $\alpha=2\pi/3$ as a function
of the annealing frequency $f$, for $\sigma=0.5$, in a system of linear size $L=32$, averaged over 
20 samples, and a coverage of $\pi/\left[3\tan\left(\pi/3\right)\right]$. 
A region with optimized structures is found at intermediate frequencies, no optimization occurs at high frequencies, and no structure
at low frequencies. Error bars are estimated from the variance of the angular distribution function in the 
bins adjacent to $\alpha=2\pi/3$ (bins are of size $\delta \alpha= 0.01$ rad). \label{fig.diagram}}
\end{center}
\end{figure}

Let us focus now on the height of the angular distribution function for $\alpha=2\pi/3$.
Figure~\ref{fig.diagram} illustrates the diagram for the ratio of the peak, $N_f(\alpha)/N(\alpha)$,  
as a function of the annealing frequency $f$, where $N_f(\alpha)$ is the value of the peak for $\alpha=2\pi/3$ and frequency $f$
and $N(\alpha)$ is the value of the same peak at the end of the initial aggregation time. As previously hypothesized (see Introduction), at high frequencies, 
the off-time is not long enough to restructure the bonds and thus the value of the peak remains the same. At low frequencies, 
once the bonds are switched off, the system is significantly randomized such that, once the bonds are switched on, there is 
not enough time to aggregate again. Note that, the initial aggregation time is much longer that the longest period considered here. 
For intermediate frequencies, the value of the peak increases by about $50\%$, suggesting that the approach to the (thermodynamic) 
honeycomb structure is improved significantly.

\begin{figure}
\begin{center}
\includegraphics[width=\columnwidth]{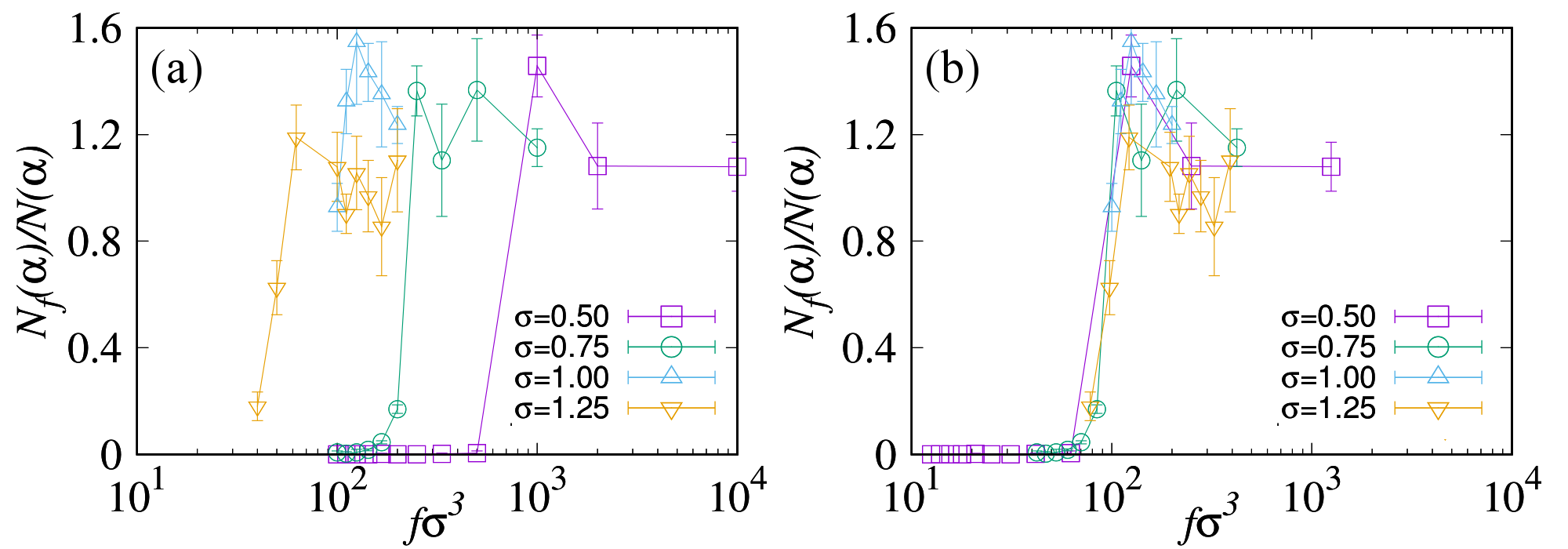}\\
\caption{(a) Angular distribution function at the honeycomb peak ($\alpha=2\pi/3$) for values of $\sigma=\{0.5, 0.75, 1, 1.25\}$ as a 
function of the annealing frequency $f$. (b) Angular distribution function at the honeycomb peak 
($\alpha=2\pi/3$) for values of $\sigma=\{0.5, 0.75, 1, 1.25\}$ as a 
function of the rescaled frequency $f\sigma^3$. Simulations were performed on a system of linear size $L=32$, averaged over
20 samples, and a coverage of $\pi/\left[3\tan\left(\pi/3\right)\right]$. Error bars are estimated from the variance of the angular distribution function in the 
bins adjacent to $\alpha=2\pi/3$ (bins are of size $\delta \alpha= 0.01$ rad). \label{fig.peaks}}
\end{center}
\end{figure}

Figure~\ref{fig.peaks}(a) shows $N_f(\alpha)/N(\alpha)$ as a function of $f$ for various values of the 
interaction range $\sigma$ and $\alpha=2\pi/3$. We find that the region where the honeycomb structure is optimized 
moved to higher frequencies when $\sigma$ increases.
Figure~\ref{fig.peaks}(b) shows the same curves, but with the frequency rescaled by $\sigma^3$, leading to a data collapse. 
This dependence is the subject of the next section.

\section{Brownian timescales}\label{sec::timescales}

\begin{figure}
\begin{center}
\includegraphics[width=0.8\columnwidth]{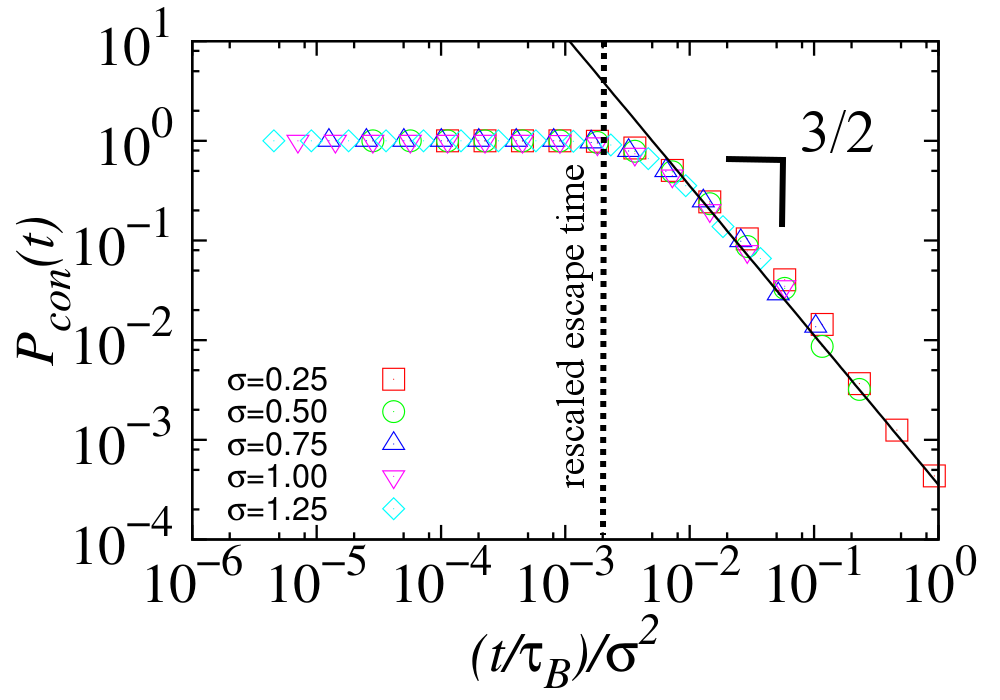}\\
\caption{Bonding probability as a function of the rescaled time $(t/\tau_B)/\sigma^2$ for two particles
initially bonded where the attractive interaction is switched off. $\tau_B$ is the Brownian time (time
necessary for a single particle to diffuse a area of $(2R)^2$, where $R$ is the particle radius) and $\sigma$ is the size of the patch.
The vertical dashed line is the rescaled escape time for particles to (first) cross the bonding threshold.
\label{fig.scaling}}
\end{center}
\end{figure}

As hypothesized, the existence of an optimal frequency results from the competition between the 
rotational and translational diffusive behavior during the off-time periods. 
When bonds are switched on again, pairs of particles will remain bonded if the centers of their patches are still at a distance less than $\sigma$.
Below we discuss how the probability of remaining bonded (bond probability) scales in time 
and the relation with the observed scaling with $\sigma^3$. 

We performed simulations for two initially bonded particles, where we 
switch off the bonds and calculate how the probability of having their patches at a distance less than $\sigma$ scales in time. 
Results for $10^4$ samples are shown in Fig.~\ref{fig.scaling}. The curves are characterized by an initial plateau and an asymptotic 
power-law decay, consistent with an exponent of $3/2$. Data for different distances $\sigma$ collapse if time is rescaled by $\sigma^2$.

When bonds are switched off, the configuration of 
individual particles (position and orientation) perform a random motion in configurational space. Thus, the absolute distance between the center of the patches 
is expected to scale as $t^{1/2}$ as in a diffusive process, i.e., time should scale as $\sigma^2$. The bond probability as defined above corresponds 
to the probability that a random walk is within a spherical region of radius $\sigma$ and the crossover is related to the escape time, 
defined as the typical time that takes a random walker to leave (for the first time) a spherical region of radius $\sigma$.

For long enough times, there is a non-zero probability that a random 
walker is found in the region of radius $\sigma$. To calculate that probability, let us consider the continuum limit of a 
random walk and define the probability distribution $P(r,\tau)$ as the probability of finding the random walk at a distance in the interval $[r,r+dr[$ 
from the origin after the rescaled time $\tau$, where $\tau=Dt$, and $D$ is the corresponding diffusion coefficient. 
In the continuum limit, this probability is the solution of the diffusion equation which, in spherical coordinates is a Gaussian distribution, 
\begin{equation}
 P(r,\tau)=\frac{1}{8(\pi\tau)^{\frac{3}{2}}}\exp{\left(-\frac{r^2}{4\tau}\right)}.
 \label{eq.gaussian}
\end{equation}
 
Defining, $P_{bond}(\sigma,\tau)$ as the probability that the random walk is within a spherical region of radius $\sigma$, then
\begin{eqnarray}
 P_{bond}(\sigma,\tau)&=&\int_0^{\sigma}drr^2P(r,\tau)4\pi= \label{eq.cumulative} \\
 &=&-\frac{\sigma}{(\pi\tau)^{\frac{1}{2}}}\exp{\left[-\left(\frac{\sigma}{2\sqrt{\tau}}\right)^2\right]}+\mathrm{erf}{\left(\frac{\sigma}{2\sqrt{\tau}}\right)}. \nonumber
\end{eqnarray}
This corresponds to the bond probability.
To obtain the long time behavior, one can expand Eq.~\ref{eq.cumulative} in powers of $\sigma/\sqrt{\tau}$, for $\sqrt{\tau}\gg\sigma$. As a result,
\begin{equation}
 P_{bond}(\sigma,\tau)\approx\frac{1}{6}\frac{\sigma^3}{\sqrt{\pi}\tau^{\frac{3}{2}}}+O\left(\frac{\sigma^4}{\tau^2}\right).
 \label{eq.final}
\end{equation}
Leading to the observed exponent of $-3/2$ and a factor of $\sigma^3$ in the optimal frequency.

\section{Final remarks}\label{sec::final}
Numerical results for annealing cycles suggest that, at low annealing frequencies, once the system is randomized, 
there is no significant time to form an ordered structure. By contrast, at high frequencies, the off-time is not long enough 
for bonds to restructure. Thus, an optimal intermediate frequency is observed. We study how this optimal frequency scales with the 
width of the interaction range (size of the patch $\sigma$). We show that the optimal frequency scales with $(\sigma/\tau)^{-3/2}$, 
where $\tau$ is the timescale defined by the Brownian process, thus related to temperature, particle shape and viscosity of the surrounding medium.

In this context, we found, not only a protocol to avoid kinetically trapped structures, but also the relevant timescales for 
patchy particles to restructure their interparticle bonds. This has potential impact on to the fabrication of novel materials with enhanced physical 
properties. For simplicity, we considered a monodisperse particle size distribution and a single type of patches. Future studies may address 
more realistic conditions.

\ack
We acknowledge financial support from the
Portuguese Foundation for Science and Technology (FCT) under Contracts
nos. EXCL/FIS-NAN/0083/2012, UID/FIS/00618/2013, IF/00255/2013, SFRH/BPD/114839/2016, and FCT/DAAD bilateral project. 

\section*{References}
\bibliography{paper}

\end{document}